\documentclass[b5paper,twoside,english]{article}
\usepackage[T1]{fontenc}
\usepackage[latin9]{inputenc}
\usepackage{amsmath}
\usepackage{amssymb}
\usepackage{graphicx}
\usepackage[authoryear]{natbib}
\usepackage{esint}
\usepackage{hyperref}



\usepackage{babel}
\begin{document}
\title{Algorithmic Composition by Autonomous Systems with Multiple Time-Scales }

\author{Risto Holopainen}
\date{ }
\maketitle

\begin{abstract}
Dynamic systems have found their use in sound synthesis as well as
score synthesis. These levels can be integrated in monolithic autonomous
systems in a novel approach to algorithmic composition that
shares certain aesthetic motivations with some work with autonomous
music systems, such as the search for emergence. We discuss various
strategies for achieving variation on multiple time-scales by using
slow-fast, hybrid dynamic systems, and statistical feedback. The ideas
are illustrated with a case study. 
\end{abstract}

\section*{Introduction}

With few exceptions, such as pure drone pieces, most music consists of auditory 
patterns that vary over time. The variation has often been found to follow 
a scale-free distribution, or a $1/f^{\beta}$
power law, which implies variation at all temporal scales and a balance
between predictability and unpredictability \citep{Levitin_rhythmspectra}.
This power law has been observed across several centuries of Western
music, and in different musical dimensions such as pitch and rhythm.
The exponent $0<\beta<2$ varies between composers and styles, where
lower values imply more variation and higher values less variation
and more predictability. 

Different temporal scales correspond to different zones of perception,
from the audio rate variation giving rise to pitch and timbral perception;
fast modulation corresponding to grittyness and roughness; slower
modulation as in tremolo and vibrato; events sufficiently brief to
process in short-term memory forming notes, phrases, gestures or motifs;
and longer processes by which we are able to segment the audio stream
into formal sections. A similar separation of structural levels can
be found in many sound synthesis languages such as Csound. There is
the low level of audio rate sound synthesis, an intermediate level
of control signals and the high level of discrete time note events.

The audio signal and its lower level attributes such as amplitude,
fundamental frequency, and spectral content are often treated as raw
material subject to external organisation by compositional procedures
or realtime input. It will be useful to distinguish two approaches
to the organisation of material, namely those of lattice-based music
on the one hand and dynamic morphology on the other \citep{Wishart_sonicart_96}.
Algorithmic composition often deals with discrete sets of pitches,
a temporal grid of onsets and duration values, and a discrete set
of symbolic dynamic levels such as \emph{pp}, \emph{p}, \emph{f},
\emph{ff}. Whereas written notes can be ordered on a lattice, dynamic
morphology is concerned with the complexity of sound objects that
change over time and cannot be easily ordered into scales. We will
outline an approach to algorithmic composition that is well adapted
to dynamic morphology, although it can also handle the discrete type
of events of lattice-based composition.

Historically, various strategies of algorithmic composition have been
applied primarily to the note level and used for score generation
\citep{Ames87}. Some notable exceptions, including the SAWDUST pieces
by Herbert Brün and the GENDYN pieces by Xenakis, using what is often
refered to as nonstandard synthesis \citep{Doebereiner2011}, manage
to bridge the separation between the low level of sound synthesis
and the higher level of large scale form. In this unified approach
to the micro and macro-levels, the waveform is composed as much as
the entire piece, sometimes applying similar procedures on all levels. 

In works inspired by cybernetics and complex systems, typically using
networks of feedback systems \citep{SanfilippoValle2013}, a side-effect
may be the upheaval of any meaningful separation of temporal scales.
Sound-generating processes that depend explicitly on a few milliseconds
of past generated sound can nevertheless result in slower large-scale
processes. Feedback systems are ubiquitus where self-organisation
and emergence are observed. 

Another feedback loop is present in virually any artistic endeavor,
namely the action -- perception -- evaluation loop. Algorithmic composition
is a circular process which may begin with the creation of an algorithm,
perhaps with a particular musical expression in mind, then the algorithm
generates a piece of music which is evaluated by listening, followed
by a cycle of further modifications of the algorithm and evaluations
of the output. The process of composition can be highly interactive
even if the interaction does not take place in realtime.

The approach outlined in this paper can be situated in the intersection
of algorithmic composition and self-organising music systems. Computer
simulations of autonomous dynamical systems require an algorithm;
hence, when applied to composition it arguably should qualify as algorithmic
composition. There are interesting parallells between this type of
algorithmic composition and (non-algorithmic) self-organising music
systems, not least from an aesthetic perspective.

Martin \citet{supper} distinguishes three categories of algorithmic
composition: 1) Modeling traditional compositional procedures, 2)
modeling original and novel compositional procedures, and 3) borrowing
algorithms from extra-musical disciplines. Style imitation by algorithmic
composition will not be our concern. As for Supper's second and third
categories, a strict separation is not necessary. Certain ideas can
be borrowed from other sciences and adapted to the needs of musical
composition. This is particularly clear when working with dynamic
systems and differential equations. When this borrowing has been established
over a few years, these previously extra-musical techniques are absorbed
into the regular toolbox of musical techniques.

Composition by algorithms or by self-organising processes is detached
from the immediate decision making typical of more intuitive approaches.
Instead of working directly with the musical material one works through
an intermediary, either by running code on a computer or by building
some electronic apparatus. The computer follows the instructions of
the program code and the electronic machinery follows the laws of
physics without the composer's direct interference. Realtime interaction
is optional, but the system might not be fully controllable. 

Autonomous systems can be formulated as a set of equations that describe
what will happen the next moment as a function of the current state.
There is no schedule or plan for future events, nor is there necessarily
any memory of past events. That is one source of difficulties often
encountered in this kind of work, as Dario Sanfilippo points out.
\begin{quote}
In my experience, the realisation of an autonomous music system which
exhibits a convincing variety and complexity over a relatively long
time span has been something difficult to achieve, even when implementing
large and articulated networks. \citep[p. 123]{Sanfilippo_2018} 
\end{quote}
Indeed, the goal is often to achieve some level of complexity and
some amount of variety over time, and the solution on offer appears
to be to add layers of mechanisms to monitor and automatically adjust
system variables.

Agostino Di Scipio's description of his work on \emph{Ecosistemico
Udibile} is illuminating in this respect \citep{DiScipio_emerg}.
The description of the piece begins with a simple electroacoustic feedback
loop from which the system grows by a process of accretion. Negative
feedback loops are added in order to balance the Larsen effect and
positive or nonlinear feedback loops are added to increase the complexity
of the system's dynamics. The system appears to have grown from inside
out like an onion, layer upon layer. When designing an autonomous
music system one cannot compose the music from beginning to end, one
has to design mechanisms that respond to situations that may arise.
To be on the safe side, one designs mechanisms also for situations
that may not arise, since what will happen in an open and highly complex
system is largely unpredictable.

In the rest of this paper we focus on deterministic autonomous dynamical
systems, beginning with an overview of previous work on ordinary differential
equations for sound synthesis. Then we discuss self-organisation and
emergence as it relates to composition with autonomous systems. Next,
we briefly consider autonomous systems, followed by an introduction
to slow-fast systems and hybrid systems that combine discrete time
and continuous flow. These concepts are then applied in a review of
previous work on feedback systems with feature extractors in the loop.
We also discuss how statistical feedback can be used as a means to
increase a system's variability. A case study shows how several of
the ideas can be applied to composition. Questions concerning the
evaluation of this class of algorithmic composition systems are addressed
in the conclusion.

\section*{Previous work on sound synthesis}
\section*{by ordinary differential equations}

Chaotic maps and ordinary differential equations (ODEs) have long been
applied to the generation of note sequences \citep{Bidlack92}. The
translation of raw data from the orbits of a dynamic system to musical
notes requires quantization into discrete values and a choice of mapping
from the state variable to musical data. Continuous time systems must
be sampled at discrete time steps, for instance by taking a Poincaré
section. Therefore, maps are inherently more suitable for the generation
of discrete events such as note sequences, whereas the smooth flow
of ODEs makes them ideal for sound synthesis. If the oscillations
are sufficiently fast the state variables may be used, after proper
amplitude scaling, as an audio stream. Slower oscillations are suitable
as modulating signals.

ODEs have not quite attained the popularity of other synthesis techniques
such as additive, subtractive, granular or physical modeling. Nonlinear
oscillators may have a non-trivial relationship between system parameters
and the qualitative character of the audio signal they produce. Pitch,
loudness and timbre may change simultaneously by the variation of
a single system parameter. Analogous codependencies exist in acoustic
instruments (e.g., the common correlation between loudness and spectral
brightness) and are not necessarily unwanted. Acoustical instruments
have been modelled with ordinary and partial differential equations.
An analysis method has been proposed that reconstructs an attractor
from a recording of an instrument tone and finds a dynamic system
capable of producing the attractor \citep{Roebel_2001}. For sound
synthesis there is no need to limit oneself to the simulation of acoustic
instruments, any dynamic system with a globally stable oscillatory
state is potentially interesting. 

Chua's circuit had been explored as a source for sound synthesis early
on \citep{Mayer-Kress_etal1993}. It was found to be capable of bassoon-like
timbres, as well as percussive sounds by using an initial transient
towards a fixed point. \citet{Rodet_Vergez_99A} were not so satisfied
with Chua's circuit on its own, but they found that extending the
system with a delay line, thereby turning it into a delay differential
equation, enriched its sonic register and provided an interesting link
to other work on physical modeling of acoustic instruments.

Before digital computers were up to the task, nonlinear differential
equations were solved on analog computers. \citet{Slater} suggested
the use of analog computers in combination with modular synthesizers
for chaotic sound synthesis. In a similarly adventurous spirit \citet{Collins_errant2008}
introduced a few unit generators for SuperCollider implementing various
nonlinear ODEs. There is an ongoing search for new chaotic systems
\citep{Sprott_2010}, many of which can be realised as electronic
circuits suitable for sound synthesis. In recent years many chaotic
oscillators have been introduced as analog modules for modular 
synthesizers\footnote{Among the notable builders of chaotic modules are Ian Fritz and Andrew
Fitch. A list of existing chaotic modules in the Eurorack format is
maintained on the Modwiggler forum: 
\url{https://www.modwiggler.com/forum/viewtopic.php?f=16\&t=152486}} and 
musicians are exploring sound synthesis using these chaotic systems.

Virtual analog modelling often needs to handle the problem of immediate
feedback paths in analog circuits, such as two mutually modulating
FM oscillators. Digital implementations usually introduce a one sample
delay for such feedback paths, but delayless, more accurate versions
can be constructed from differential equation models of the original
system \citep{Medine_2015}. \citet{stefanakis_synthesis_ODE} introduced
a few useful techniques by relating complex-valued, time-dependent
systems of ODEs with input signals to more familiar concepts of sound
synthesis and filtering. Complex variables have the advantage that
amplitude and frequency can be modelled in a single variable. Using
noise as input, these systems become stochastic differential equations.

\citet{Jacobs_diffeq} describes a system of connected Fitzhugh-Nagumo
oscillators on a graph, which are excited by a wave modelled by a
partial differential equation. The function of the travelling wave
is similar to a higher level control function, and Jacobs describes
it as a sequencer or a rudimentary tool for algorithmic composition.
This integration of sound synthesis and control level signals is somewhat
similar to the approach that we will pursue in this paper.

Although far from complete, this literature survey hopefully shows
the diversity of approaches to differential equations in the synthesis
of musical signals. Ordinary, as well as delay, stochastic, and partial
differential equations have been explored and may be useful as sources
of variation on multiple time scales. 

\section*{Emergence and surprise}

An important motivation for making music with autonomous systems is
the search for self-organisation and emergence. A useful summary is
provided by \citet{wolf_holvoet}, who list a few criteria often considered
crucial for emergence:
\begin{enumerate}
\item Global behaviour, properties, patterns, or structures result from
interactions at a lower level. These higher level phenomena are called
\textquotedblleft emergents\textquotedblright .
\item The global phenomenon is novel and not reducible to the micro-level
parts of the system. In the words of Wolf and Holvoet, \textquotedblleft radical
novelty arises because the collective behaviour is not readily understood
from the behaviour of the parts.\textquotedblright{} We return to
this point below.
\item Emergents maintain a sense of identity or coherence over time.
\item For emergence to occur, the parts need to interact. Therefore,
the system should be highly connected at the low level.
\item Decentralised control of the system implies that no single part is
responsible for the macro-level behaviour; the system as a whole is
not controllable.
\item In turn, the decentralised structure makes the system flexible and
robust against small perturbations. Parts of it may be replaced without
changing the emergent.
\end{enumerate}
Self-organisation, according to a view that goes back to Ashby, occurs
when the degree of organisation or order increases within a system
as it evolves by its own dynamics. By this understanding, any dissipative
dynamic system that approaches an attractor is a self-organising system. 

Some rigorous and quantifiable approaches to emergence and self-organisation
have been proposed \citep{Prokopenko_etal}. A set of measures introduced
by \citet{Gershenson_Fernandez} relate the amount of information
or Shannon entropy at the input to that at the output of a system.
Although it is not always clear how the amount of information should
be measured, in particular when considering the musical output of
a complex system where perceptual criteria should arguably play a
decisive role, the idea of comparing input and output information
is worth considering. We will return to this point in the Conclusion.

In an overview of some interfaces for self-organising music, \citet{Kollias2018}
emphasises electro-acoustic feedback as a primordial element around
which many of the works have been structured. Although the openness
to the acoustic environment perhaps puts these systems in a special
category, feedback can be explored in the digital or analog domain
as well, or in any mixture of domains. An emerging category that fits
the description of self-organising music interfaces very well is modular
synthesizers and what is often refered to as ``self generating patches''\footnote{For sound examples, 
patch ideas and general discussion, see the thread
\textquotedblleft Self generating patches....tips and ideas ?\textquotedblright{}
started at the Modwiggler forum on March 24, 2011: 
\url{https://www.modwiggler.com/forum/viewtopic.php?f=4\&t=31698}}. 
These are analog or hybrid analog/digital systems set up in large
networks of modules that may run autonomously and produce complex sequences of music. 

Surprise, as well as emergence, are frequently mentioned as desired
qualities in work involving self-organising music systems. Emergence
is often described in terms of the expectancies of an observer, defining
\textquotedblleft the quality of unexpectedness of the results\textquotedblright{}
\citep[p. 18]{SanfilippoValle2013}; see also the already quoted view
that novelty arises because collective behaviour is not readily understood
from the behaviour of the parts \citep{wolf_holvoet}. This would
seem to imply that emergence is a mere side-effect of not knowing
exactly what to expect, of lacking a full understanding of the system's
dynamics. Yet, one can argue, having an inkling of what the system
is capable of is necessary for building up an expectation -- that
can then be thwarted when the system behaves in an unexpected way.
In any case, as \citet{Kivelson} point out, the definition of emergence
according to which \textquotedblleft something is qualitatively new
if it cannot be straightforwardly understood in terms of known properties
of the constituents\textquotedblright{} suffers from many shortcomings;
\textquotedblleft perhaps the most glaring is that it implies that
as soon as something is understood it ceases to be emergent.\textquotedblright{} 

Novelty effects also wear out with repetition, which shows the troubling
ephemerality of emergence as defined in terms of the observer's reactions.
Nevertheless, anticipation of what will happen next in a piece of
music is a crucial part of the listening experience, as \citet{Huron2007}
discusses at length. From an evolutionary perspective, surprise indicates
a failure to predict an event in our environment, which ultimately
can be bad for our prospects of survival. As Huron points out, we
actually enjoy being right in our predictions of what is going to
happen next in a piece of music, including correctly predicting the
regular recurrence of a downbeat or the chord sequence of a cadence.
This seemingly contradicts the common wisdom that we enjoy surprises
in music. Violated expectations, after all, produce reactions like
frisson, awe, or laughter.

The surprise of a practitioner of self-organising music as the system,
for some poorly understood reason, generates an output that is more
complex than expected is very different from the reaction of a listener
who is not aware of what is going on inside the system. It is by no
means illegitimate to seek out these surprising situations as a practitioner,
but we should be aware that for the uninformed listener, the surprise
is a function of previous listening experiences and whatever expectations
the piece itself sets up, in contrast to the expectations of the composer,
the one who built the system and knows a few things about its inner
workings.

As for emergence, \textquotedblleft musical form emerges from interactions
composed at the signal level\textquotedblright{} as \citet{DiScipio_emerg}
puts it concerning his own work. Indeed, this would be a good example
of emergence independent of the observer and their reactions.

\section*{Autonomous systems}

We now turn to a more technical description of some aspects of dynamic
systems that are of importance in multi-scale algorithmic composition.
The term 'autonomous system' has a rather precise meaning in the context
of dynamic systems, and a less stringently defined meaning in the
context of algorithmic or generative music.

Let us recall the definition of a dynamic system in continuous time,
\emph{t}, with a state variable $x\in\mathbb{R}^{n}.$ A general ODE
is described by an equation

\[
\dot{x}\equiv\frac{dx}{dt}=f(x;p(t))
\]
evolving from an initial condition $x(0)=x_{0}$ with a constant or
time-variable parameter $p(t)$. An autonomous system has no explicit
time dependence, so it has the form $\dot{x}=f(x;p)$ where \emph{p}
is constant. 

From a musician's point of view, autonomy rules out realtime interaction
and external control of the system. Thus, autonomous systems
have no place, without abuse of terminology, in interactive live-electronics
where system parameters or the state variable itself may be put under
the performer's influence, nor can the system receive an input signal.

The opposite of an autonomous system is a forced or driven system.
To complicate things, we note that the distinction is also a matter
of perspective. 

Consider the equation $\dot{x}=-x+\sin t$ which has the non-autonomous
forcing term $\sin t$. This system can be reformulated as an autonomous
system in two ways. First, one could introduce a new time variable, $\tau,$
and write the system as

\begin{eqnarray*}
\dot{x} & = & -x+\sin\tau\\
\dot{\tau} & = & 1.
\end{eqnarray*}
Alternatively, $\sin t$ can be expressed as the orbit of an harmonic
oscillator which takes two new state space variables and initial conditions. 

Similar distinctions can be discussed in the context of self-generating
patches on modular synthesizers. The goal is to build a patch that
generates interesting musical sequences of its own accord without
manual interference. Is the patch truly autonomous (\textquotedblleft self-generating\textquotedblright{}
or \emph{autopoietic}) if there is a sequencer driving it, or an LFO
or noise source? One could argue that the patch is \emph{more} autonomous---a
matter of degrees---the fewer sources of modulation or discrete events
there are that are not themselves modulated by other parts of the
patch.

Interconnectedness may therefore be a more useful criterion than autonomy.
In fully connected systems all parts receive input from all other
parts. In the limit everything is bidirectionally coupled to everything
else. Then there can be no external sources of modulation or control;
hence, the system must be autonomous. Notice also that interconnectedness
was listed as one of the criteria for emergence (see point 4 in the
list in the previous section).

\section*{Slow-fast systems }

Relaxation oscillators such as the van der Pol oscillator or the stick-slip
mechanism in bowed string instruments are well-known examples of slow-fast
systems. A number of special techniques have been introduced to simplify
the treatment of such multi-scale systems (see \citet{Strogatz94}
for some of them). Our motivation for discussing slow-fast systems
is that they provide a convenient conceptual framework for describing
compositional models that integrate audio synthesis and larger scale
levels.

A general slow-fast system in two time-scales may be written

\begin{alignat}{2}
\dot{x} & = & \epsilon f(x,y)\label{eq:slowfast}\\
\dot{y} & = & g(x,y)\nonumber 
\end{alignat}
where $\epsilon$ is a small positive time scaling factor, $x\in\mathbb{R}^{m}$
is the slow subspace and $y\in\mathbb{R}^{n}$ is the fast subspace.
For example, an audio rate oscillator and an LFO modulating each other
could be modeled as a slow-fast system. With mutual coupling between
the slow and fast subspaces their dynamics are intertwined, although
with a sufficient separation of time-scales or a loose enough coupling
some simplifying assumptions can be made for a qualitative understanding
of the dynamics of each subspace. 

In particular, in the fast subspace $\dot{y}=g(x,y)$ the variable
\emph{x} may be regarded as a set of slowly drifting parameters. As
the parameters drift by some small amount the fast subsystem may change
smoothly. The drifting parameters may also cause a bifurcation producing
a qualitatively different dynamics in the fast subsystem. As long
as the fast system does not bifurcate one could think of it as an
attractor continuously varying in shape and the fast orbit as permanently
being in a transient state chasing the current attractor \citep{Ruelle87}.

Conversely, with the slow subsystem modulated by the fast subsystem
the rapid oscillations may have an effect similar to noise, where
the average position $\langle y\rangle$ in the fast system's phase
space will act like a constant parameter value with an added \textquotedblright noise\textquotedblright{}
term $\xi$. Then we may write the slow system as 

\[
\dot{x}=\epsilon f(x,\langle y\rangle+\xi).
\]
The usefulness of these simplifications depends on the strength of
the coupling between the two subsystems, the separation of their characteristic
time-scales and the exact form of the equations. 

\citet{Fujimoto_Kaneko} have shown that in a chain of coupled chaotic
systems, each slower than the next one by a constant factor, under
certain conditions the fastest system can influence the slowest. For
this to happen the separation of time-scales must be in a certain
range; there must be a bifurcation in the fastest system and the bifurcation
must cascade through to the slower systems. In their particular model,
despite mutual coupling, the influence went only from faster to slower
systems. It is probably more common to have slow subsystems influence
the faster subsystems.

Slow-fast systems commonly describe spiking or bursting dynamics,
such as firing neurons, where the state variable moves slowly for
most of the time and then jumps or oscillates rapidly. In contrast,
the slow-fast systems we are interested in should have a more or less
permanent fast time-scale for audio signals and slower time-scales
for synthesis parameters to ensure variation over time.

\section*{Hybrid systems}

Other useful ideas are differential equations with discontinuous right-hand
side and hybrid systems that combine the continuous flow with the
discrete time of maps. Many theoretical results about flows assume
smooth vector fields, but non-smooth systems such as piecewise smooth
functions are useful models of many mechanical and electronic phenomena.
In particular, what makes this class of systems interesting is that
they allow for sudden changes.

Hybrid systems can have their discontinuities imposed at certain points
in time, such as

\[
\dot{x}=f_{i}(x)\quad\textrm{for}\,t\in T_{i},\:i=1,2,\ldots,N
\]
where $T_{i}$ are disjoint sets of time intervals whose union covers
all time for which the system is defined, such that the flow follows
different equations at different time intervals. Alternatively, the
system can switch between different sets of continuous flow equations
when the state variable passes from one region of the state space
to another. Hysteretic switching is also possible, where the switching
happens only if the state variable approaches the switching point
from a certain direction \citep{Saito_switched}. 

Since we are interested in autonomous systems we will consider discontinuities
of the form

\[
\dot{x}=\begin{cases}
f_{1}(x), & x\in\Omega_{1}\\
f_{2}(x), & x\in\Omega_{2}
\end{cases}
\]
induced by switchings depending on position in state space. For simplicity
we consider a system separated into two distinct regions, $\Omega_{1,2}$,
but the idea easily extends to any number of regions.

\begin{figure}
\includegraphics[width=1\textwidth]{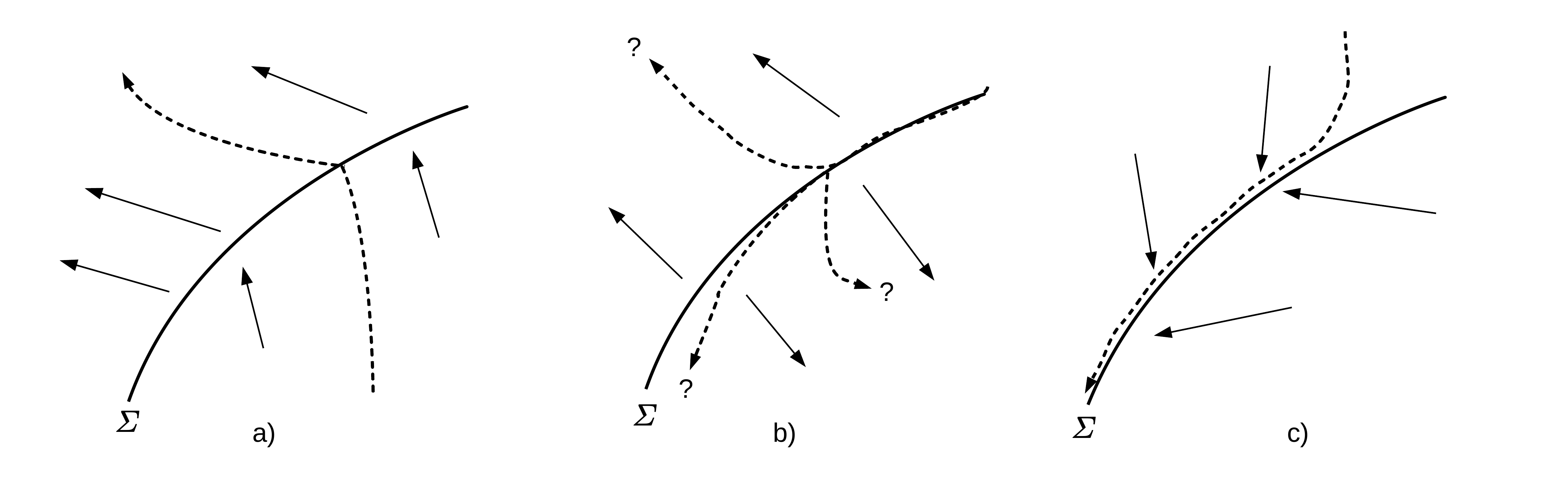}

\caption{Dynamics near the switching manifold. \emph{a}) The vector fields
on both sides point in roughly the same direction and the flow
crosses the switching manifold. \emph{b}) If the vector fields point
away from the switching manifold the solution might not be unique.
\emph{c}) When the vector fields point inwards to $\mathit{\Sigma}$
the trajectory may stick or slide along it.\label{fig:switch}}
\end{figure}

Suppose each region $\Omega_{i}$ of continuous flow is governed by
the equation $\dot{x}=f_{i}(x)$. The border between regions is called
the switching manifold, $\mathit{\Sigma}$. What happens when $x\in\mathit{\Sigma}$
is not obvious, the system might not even have a unique solution \citep{danca_uniqueness_discontinuous_ode}.
Three different situations near a switching manifold are illustrated
in Figure \ref{fig:switch}. When the vector fields on one side point
into the other side and the vector field on the other side points
further away (\emph{a}) the trajectory will pass through. In the case
of vector fields pointing away from the switching manifold there may
be no unique solution for an orbit starting on $\mathit{\Sigma}$
itself (\emph{b}), and finally (\emph{c}), if the vector fields on
both sides point inwards to $\mathit{\Sigma}$ the trajectory may
approach it and start sliding along it. 

In addition to the familiar bifurcations that occur in smooth systems
some new bifurcation scenarios are only observed in discontinuous
or nonsmooth systems \citep{Makarenkov_Lamb}. So called \emph{grazing}
happens when a periodic orbit touches the switching manifold, which
models situations such as a swinging clapper just touching a church
bell. Another example is the friction causing the squealing noise
of brakes. 

Modular synthesizers are essentially hybrid systems. Oscillators and
filters generate and process continuous time signals, whereas triggers
and clock signals are discrete time events. Similarly to the switching
functions discussed above, the continuous signals can be segmented
with sample \& hold or analog shift registers.

In autonomous ODEs used for algorithmic composition, hybrid systems
allow for smooth flows suitable for audio signals to have points of
instant change. Applied to frequency, one can articulate a discrete
set of pitches instead of having a constant glissando. Combining a
slow-fast system with switching is particularly interesting when the
discontinuities are defined on the slow subspace, as we will demonstrate
in the case study below. 

\section*{Feedback from feature extractors}

In the search for mechanisms that equilibrate a feedback system it
is quite natural to turn to feature extractors, such as in Di Scipio's
\emph{Feedback Study} \citep{DiScipio_emerg}. My previous research
centred on a class of autonomous systems comprised of an oscillator
or signal generator, a feature extractor analysing the oscillator's
output, and a mapping unit that transforms the feature extractor's
output to synthesis parameters for the signal generator \citep{Holopainen_PhD2012};
I have proposed to call these systems Feature Extractor Feedback Systems,
or FEFS for short.

For a feature extractor to be useful inside a feedback system, it
should process short segments of the most recent audio output and
be able to provide output without too much latency. Time domain feature
extractors that generate audio rate output are particularly suitable
for this purpose. Block-based processing using DFT or other transforms
typically deliver output values at a much slower rate. Unless the
output is interpolated, block-based processing imposes its own regular
pace of updates which tends to become a dominant and easily audible
effect.

By adapting to their own output FEFS have useful applications such
as automatic pitch correction of nonlinear oscillators with unknown
functional relations between parameters and pitch. Another interesting
scenario would be to connect several FEFS in networks where each unit
analyses and responds to the other units.

A few detailed studies of various FEFS models led me to conclude that
the feature extractor's most salient contribution was a smoothing
effect, which we will explain shortly. A broad class of FEFS can be
described by the equations

\begin{alignat*}{1}
x_{n+1} & =g(\theta_{n})\\
\theta_{n+1} & =f(\theta_{n},\pi_{n})\\
\pi_{n+1} & =m(\phi_{n})\\
\phi_{n+1} & =a(x_{n},x_{n-1},\ldots,x_{n-L+1})
\end{alignat*}
where $x_{n}$ is the audio output, $\theta_{n}$ is an internal state
or phase variable of the signal generator, $\pi_{n}$ are the synthesis
parameters, and $\phi_{n}$ is the output of a feature extractor which
operates on the last \emph{L} output samples. All variables may be
vector valued.

Let us consider feature extraction using zero crossing rate (zcr)
as an example. The zero crossing rate can serve as a crude pitch estimator
or a descriptor of spectral balance. There is a well-known trade-off
between temporal acuity and precision; a longer feature extractor
window provides more accurate frequency estimates but also smears out sudden
changes in the analysed signal, whereas shorter windows respond faster
to changes but with less accuracy.

There are a few different implementations of zcr, a popular choice
being to count the number of zero crossings during the past \emph{L}
samples and divide by \emph{L}. Another, probably less common way
is to tally up to a certain fixed number of \emph{N} zero crossings
and then divide by the number of samples elapsed since the first counted
crossing. This method adapts to the signal's content and uses a longer
effective window length for lower frequencies than for high frequencies. 

The mapping $\pi_{n+1}=m(\phi_{n})$ from feature extractors to synthesis
parameters also plays an important role. Making this mapping highly
nonlinear increases the susceptibility of the FEFS to exhibit wild
behaviour, whereas smoother mappings are likely to make $\phi_{n}$
and $\pi_{n}$ settle on some fixed values, thus causing the generator
to output a signal with constant synthesis parameters. As a rule of
thumb, increasing the feature extractor's analysis window length has the 
opposite effect of increasing the nonlinearity of the mapping. 

In order to emphasise an essential point of FEFS we may grossly simplify
and reformulate the system as a \emph{filtered map}. The feature extractor
operates on a running block of samples, and its effect can be modelled
as a time average combined with some nonlinearity. The simplified
system

\begin{alignat*}{2}
x_{n+1} & = & f(x_{n},y_{n})\\
y_{n} & = & \langle g(x_{n})\rangle
\end{alignat*}
lumps together the oscillator and mapping in $f(x,~y)$, and \emph{y}
represents the output of the feature extractor; $g(x)$ represents
some nonlinear function of the audio signal, whose time average $\text{\ensuremath{\langle x\rangle}}$
is taken over the last \emph{L} samples, corresponding to the effective
window lenght of the feature extractor. 

Clearly, the averaging smooths out any rapid changes, so \emph{y}
is a slow variable. Notice that even if $f(x,~y)$ were a chaotic
map, a long enough smoothing of its orbit in the feature extractor's
averaging part makes it likely that the parameter \emph{y} will approach
a constant value with perhaps some small fluctuations. In other words,
this situation corresponds exactly to the slow-fast system (Eq. \ref{eq:slowfast})
where we approximated the slow variable as a constant plus noise term.

It has proven difficult to design FEFS that exhibit non-trivial behaviour
over extended time. Often there is a more or less prolonged initial
transient phase after which the slow variables ($\phi_{n}$ and $\pi_{n}$)
approach some fixed state. The analysis in terms of a slow-fast system
explains why this often happens. Chaotic solutions are also attainable
with short effective window lenghts and strongly nonlinear mapping
functions. 

\section*{Statistical feedback and mapping with memory}

Deterministic FEFS, as formulated above, are limited by the fact that
the mapping from feature extractors to synthesis parameters involves
no memory of its past beyond the effective lenght of the feature extractor.
No monitoring mechanism allows the system to discover if it has got
stuck on some repeating cycle that extends beyond the length of the
feature extractor. It might resemble an improviser suffering from
a loss of short-term memory, who would not be able to guide the performance
in any particular direction other than wherever the haphazard steps
of a Markov chain brings it. 

Algorithmic compositions using autonomous systems may stray far from
the syntax and material of the common-practice music that has been
extensively studied by music scholars. Nevertheless, certain insights
from music theory and music psychology may be relevant, even though
its focus has been lattice-based structures rather than dynamic morphology,
to borrow Wishart's terminology again. Research in music psychology has
provided ample evidence that statistical learning of musical patterns
plays an important part in forming listener expectations \citep{Huron2007}.
First order probabilities, such as the distribution of pitches or
scale degrees, allow us to identify scales and their tonal centres.
Higher order probabilities, including transition probabilities between
pitches or durations, contribute to the recognition of different styles.

As noted in the introduction, it is a common experience that the goal
of variety over time may be elusive in autonomous music systems. Variety
can always be increased by adding another regulatory mechanism, a
new layer of the onion that makes up the algorithm. A strikingly simple
and efficient technique called dissonant counterpoint was pioneered
by James Tenney. Originally it was used to enforce a certain amount
of variation on randomly generated pitch sequences (or dynamics, durations,
whatever parameter one wishes to control). Random sequences may have
occurrences of short subsequences that appear more orderly than one
might naively expect, such as immediate repetitions or alternations
between elements. Tenney's method guarantees that such close repetitions
are ruled out. Simply put, Tenney's scheme goes as follows \citep{Polansky_2011Tenney}:
\begin{enumerate}
\item Initialise an array of \emph{N} entries all with the same positive
number (say, 1). Each element corresponds to a unique pitch. 
\item Interpret the values stored in the array as relative probabilities.
Pick an element randomly using its relative probability.
\item Reset the value of the chosen entry to zero and increase the values
of all other entries.
\item Repeat from step 2.
\end{enumerate}
This simple scheme will make a direct repetition of a pitch impossible
and a close repetition unlikely. Tenney's method cannot be used as
such in a deterministic autonomous system since it relies on random
choices, but the idea of keeping track of the statistics of past states
can be generalised to suit our needs.

In a FEFS, one can keep track of the relative frequency of past synthesis
parameter or feature extractor values. In practice, one would use
a number of discrete bins for a histogram and introduce a mechanism
tipping the system in a direction that favours the production of less
often occuring values. This, in effect, expands the dynamic system's
memory of its past behaviour without neccessitating storage of the
entire time series. There is no guarantee that the system will not
settle on cyclical patterns, but at least the cycles are likely to
be longer than they would otherwise be (see \citet[pp. 275-280]{Holopainen_PhD2012}
for further details). 

The goal distribution of synthesis parameter values does not have
to be uniform. Uneven distributions lead to a differentiation of common
and rare events. If these events are perceptually distinguishable,
the rare ones are those that are prone to carry some significance
for the listener by being outliers or exceptions. Statistical learning
is at play also in the course of listening to a piece for the first
time (this is what \citet{Huron2007} calls dynamic expectations,
in contrast to those expectations that pertain to a whole genre which
he refers to as schematic). Expectations are shaped by the probability
distribution of events as they occur in a piece. 

The information content and information density of pieces of music (amount of
information over time) has been studied in music theory for decades.
The less probable an event is, the higher its information content.
As \citet{Temperley2019} points out, it would be a formidable task
to quantify the amount of information in any musical piece, since
everything (melodic pitches, harmony, rhythm, timbre, and other aspects)
contributes to information. And this is still only considering common-practice
music with large corpora available for study and the information-carrying
units relatively easy to identify. 

Algorithmic composition beginning from sound synthesis, where higher
levels emerge from processes at lower levels, poses the additional
difficulty of relating synthesis parameters to resulting audio signals,
and the audio signals to their perceptual correlates, and of defining
the units that carry information. There is a gap between theories
of information applicable to the symbolic note level and the dynamic
morphology articulated by sound synthesis with freely flowing synthesis
parameters. Some possible approaches to information density or complexity
will be discussed in the conclusion. 

\section*{Case study: The auto-detuning system}

We turn now to a simple three-dimensional ODE which serves as a building
block in two algorithmic compositions. It is an example of a slow-fast
system with discontinuous derivative on its right-hand side. The system
consists of two oscillators that are detuned by an amount that depends
on the amplidude of the sum of the oscillators.

In its simplest form, the system is given by 

\begin{eqnarray}
\dot{\theta}_{1} & = & \omega+\delta\nonumber \\
\dot{\theta}_{2} & = & \omega-\delta\label{eq:detuned}\\
\dot{\delta} & = & c\left|\sin\theta_{1}+\sin\theta_{2}\right|-\delta\nonumber 
\end{eqnarray}
with a detuning parameter $\delta\geq0$, phase variables $\theta_{1,2}$
and frequency $\omega$. The parameter $c\geq0$ indirectly affects
the amount of detuning. 

For now, let $\omega=1$ and let $\Omega_{i}=1\pm\delta$ be the frequencies
of the two oscillators. If the frequency were stable the two oscillators
would be tuned to a constant ratio $R=\Omega_{1}/\Omega_{2}$ which
is known as the rotation number, and may be calculated as

\[
R=\underset{T\to\infty}{\lim}\frac{1}{T}\intop_{0}^{T}\frac{\theta_{1}}{\theta_{2}}dt.
\]
For $0<c<1/4$ the two oscillators are perfectly locked in sync, so
$R=1$. As \emph{c} increases from \emph{$c=1/4$} the rotation number
decreases. 

Assuming that $\delta$ approaches some constant value, the phases
will increase at a constant rate. Under such conditions we can calculate
an average value of the expression in the third equation of Eq (\ref{eq:detuned}),
for which we introduce the variable 

\[
\vartheta=\left|\sin\theta_{1}+\sin\theta_{2}\right|.
\]
If we knew the time average $\langle\vartheta\rangle$ we could solve
the equation 

\[
\dot{\delta}=c\langle\vartheta\rangle-\delta
\]
separately, which is easy; it will simply approach $c\langle\vartheta\rangle$
asymptotically. Under certain assumptions ($\delta$ being constant
and \emph{R} irrational), the average can be found by evaluating

\[
\langle\vartheta\rangle=\frac{1}{(2\pi)^{2}}\iintop_{S}\left|\sin\theta_{1}\sin\theta_{2}\right|d\theta_{1}d\theta_{2}
\]
over the region $S=[0,2\pi]\times[0,2\pi]$, which turns out to be
$\langle\vartheta\rangle=8/\pi^{2}\approx0.81$.

\begin{figure}
\includegraphics[width=1\textwidth]{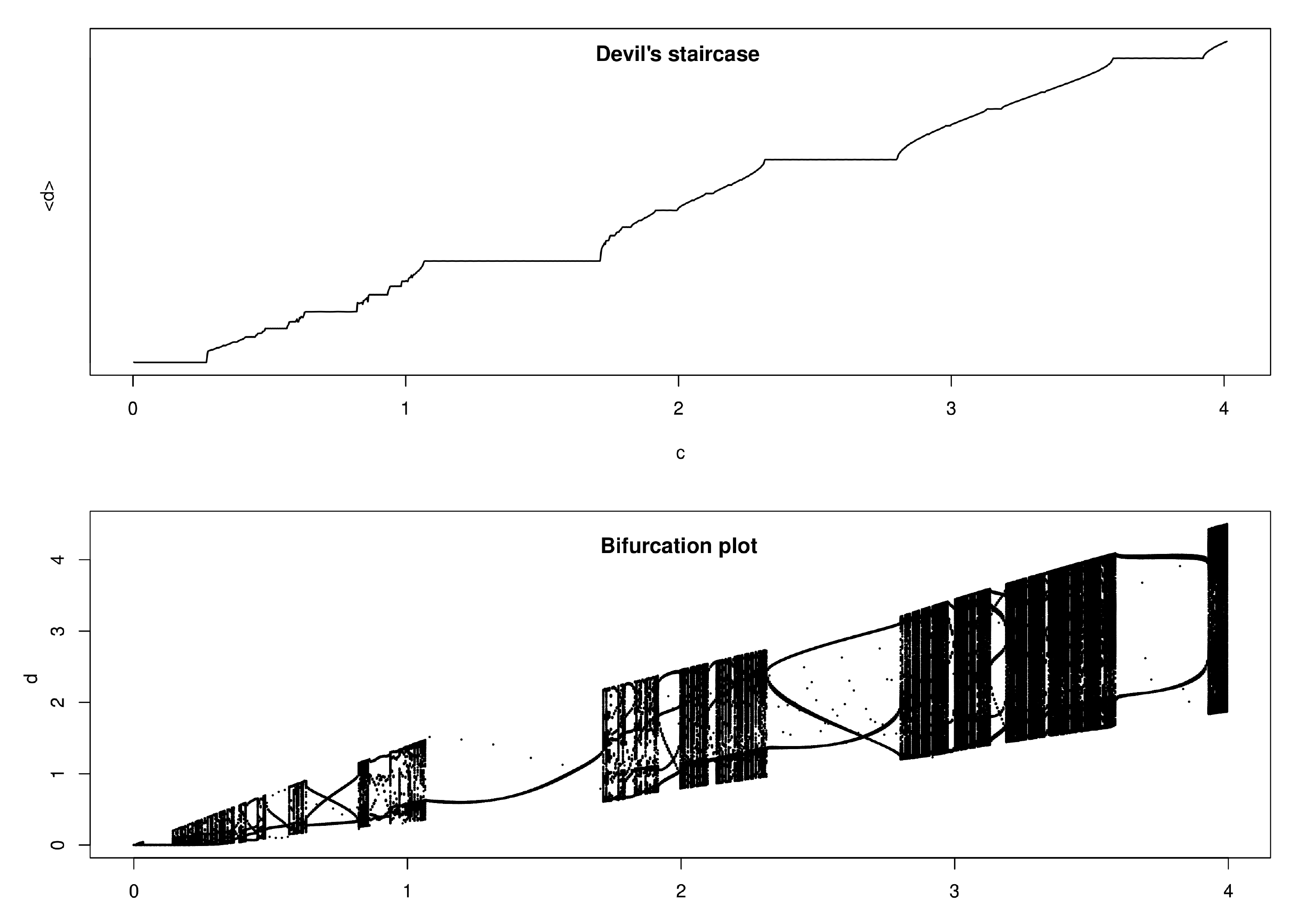}

\caption{Top: A devil's staircase appears when plotting the time average of
$\delta$ against \emph{c}. Bottom: Poincaré section ofover the same
parameter range. \label{fig:devils_staircase}}

\end{figure}

Although $\delta$ is not constant but fluctuates over time, it is
revealing to plot its average value as a function of \emph{c} (Fig.
\ref{fig:devils_staircase}). Doing so, we find the familiar fractal
graph of the devil's staircase, implying that for certain intervals
of \emph{c} the average $\langle\delta\rangle$ locks to a constant
value.

Let us see how this 3D ODE resembles a FEFS. The two oscillators make
up the signal generator; there is a rudimentary feature extractor
(taking the absolute value of the oscillators' sum and lowpass filtering
works as an envelope follower); and the estimated amplitude envelope
is mapped to the detuning synthesis parameter. In terms of slow-fast
systems, the oscillators are the fast variables (the first one faster
than the second), and the detuning is the slow variable.

The auto-detuning system is deeply embedded within a program that
generates a piece called \emph{Auto-detune}. As used there, it has
been bestowed with a few additional time-varying parameters wich in
turn are parts of other regulating mechanisms; there are also several
instances of the system as well as other systems all connected in
a complicated web. The system then takes the form

\begin{eqnarray*}
\dot{\theta}_{i} & = & \omega_{t}\pm\delta\\
\dot{\delta} & = & c_{t}\left|\sin\frac{\theta_{1}}{K_{1}}+\sin\frac{\theta_{2}}{K_{2}}\right|-\delta
\end{eqnarray*}
with certain functions doing the updating of all time-varying parameters.
Notice that division by $K_{i}\gg1$ is a way to transform the phases
originally used for audio signals into much slower variables. 

Another variant of the auto-detuning system is used as part of a larger
autonomous system in the piece \emph{Megaphone}. The piece unfolds
as a series of beating sinusoids fading in and out, sometimes making
jumps in pitch. Several trials while tuning the system parameters
resulted in processes that produced variation for a few minutes and
then approached an equilibrium state. 

\emph{Megaphone} combines ideas from the previous piece \emph{Bourgillator}\footnote{See \url{https://ristoid.net/research/bourgillator.html}}
and the above described detuning system. \emph{Bourgillator} consists of
a network of oscillators whose frequencies are updated by a function
of the output amplitudes of the oscillators. Specifically, the ordering
of their relative amplitudes determines the frequency of each oscillator.
The oscillators are also phase coupled as in the Kuramoto system
\citep[e.g.][]{Strogatz_2000}, which allows for synchronisation of
all oscillators or clusters of oscillators. 

\emph{Megaphone} is built on a larger system of coupled subsystems, and is
roughly described by the following set of equations. First there is
a modified auto-detuning part,

\begin{eqnarray*}
\dot{\theta}_{i} & = & \omega_{i}+\delta_{i}-\kappa\sin(\theta_{i}-\varphi_{j})\\
\dot{\varphi_{i}} & = & \omega_{i}-\delta_{i}-\kappa\sin(\varphi_{i}-\theta_{j})\\
\dot{\delta}_{i} & = & c_{i}\left|u_{i}+v_{i}\right|-\delta_{i}
\end{eqnarray*}
where $i=1,2,...,N,j=i+3\;(\bmod N)$, with $N=13$ oscillators used in the realisation 
of the piece. The auxilliary variables $u, v$ are defined by

\begin{eqnarray}
u_{i} & = & \alpha_{i}\sin(\theta_{i}/M_{u}) \nonumber \\
v_{i} & = & \alpha_{i}\sin(\varphi_{i}/M_{v}). \label{eq:mumu}
\end{eqnarray}

The main difference from the bare-bones detuning system (Eq. \ref{eq:detuned})
is the phase coupling and the amplitude variables $\alpha_{i}$.
Next, we introduce a slow variable in the form of an envelope follower
applied to the oscillators' outputs. A simple envelope follower tracing
the amplitude of a faster variable $u(t)$ can be realized with the
equation 

\[
\dot{A}=\tau(u^{2}-A)
\]
using a time scaling constant $0<\tau\ll1$. The envelope followers
are applied also to the variables $v_{i}$. Then the amplitude envelopes
are used as inputs to a function $g({\bf A})$, 

\begin{equation}
g_{i}({\bf A})=\sum_{j}\beta_{j}U(A_{i}-A_{i+j}),\label{eq:g_func}
\end{equation}
where $U$ is the Heaviside step function and $\beta_{j}>0$ is a
decreasing sequence of coefficients, the output of which sets the
oscillator frequencies to

\[
\omega_{i}=g_{i}({\bf A}).
\]
The step function of course is discontinuous, and so is the sum of
step functions in Eq. (\ref{eq:g_func}). Since the envelopes are
slow variables the output of $g({\bf A})$ can be expected to remain
constant for certain intervals of time and then jump to another value.
Thus, the function $g({\bf A})$ produces a stepped sequence of pitches.
As soon as one of the envelopes overtakes another envelope in amplitude
the function will produce a new pitch for at least one of the oscillators. 

Now, the goal is to keep these pitch changes happening. The system should
be designed such that the amplitude envelopes do not immediately settle on
a fixed order of relative loudness, because when that happens the piece
no longer evolves, it has frozen into a final fixed state. Obviously an initial
condition must be chosen so as to avoid landing directly on such a steady state.
A trick to introduce some more variability is to inject a small amount of
neighboring envelopes into each envelope follower,

\[
\dot{A}_{i}=\tau(u_{i}^{2}+v_{i}^{2}-\gamma_{1}A_{i}+\gamma_{2}A_{j})
\]
and also modifying the function $g({\bf A})$ to compare the sizes
of amplitude \emph{differences} between pairs of envelopes instead
of the envelopes by themselves.

Next we introduce a set of even slower variables

\[
\dot{\psi}_{i}=A_{i}-k\psi_{i}
\]
and use them to update the oscillator amplitudes 

\[
\alpha_{i}=\cos(\psi_{i})
\]
as well as other parameters such as $M_u$ and $M_v$ as defined in Eq. (\ref{eq:mumu}) 
that are supposed to change at a slow pace.

\begin{figure}
\begin{centering}
\includegraphics[width=0.95\textwidth]{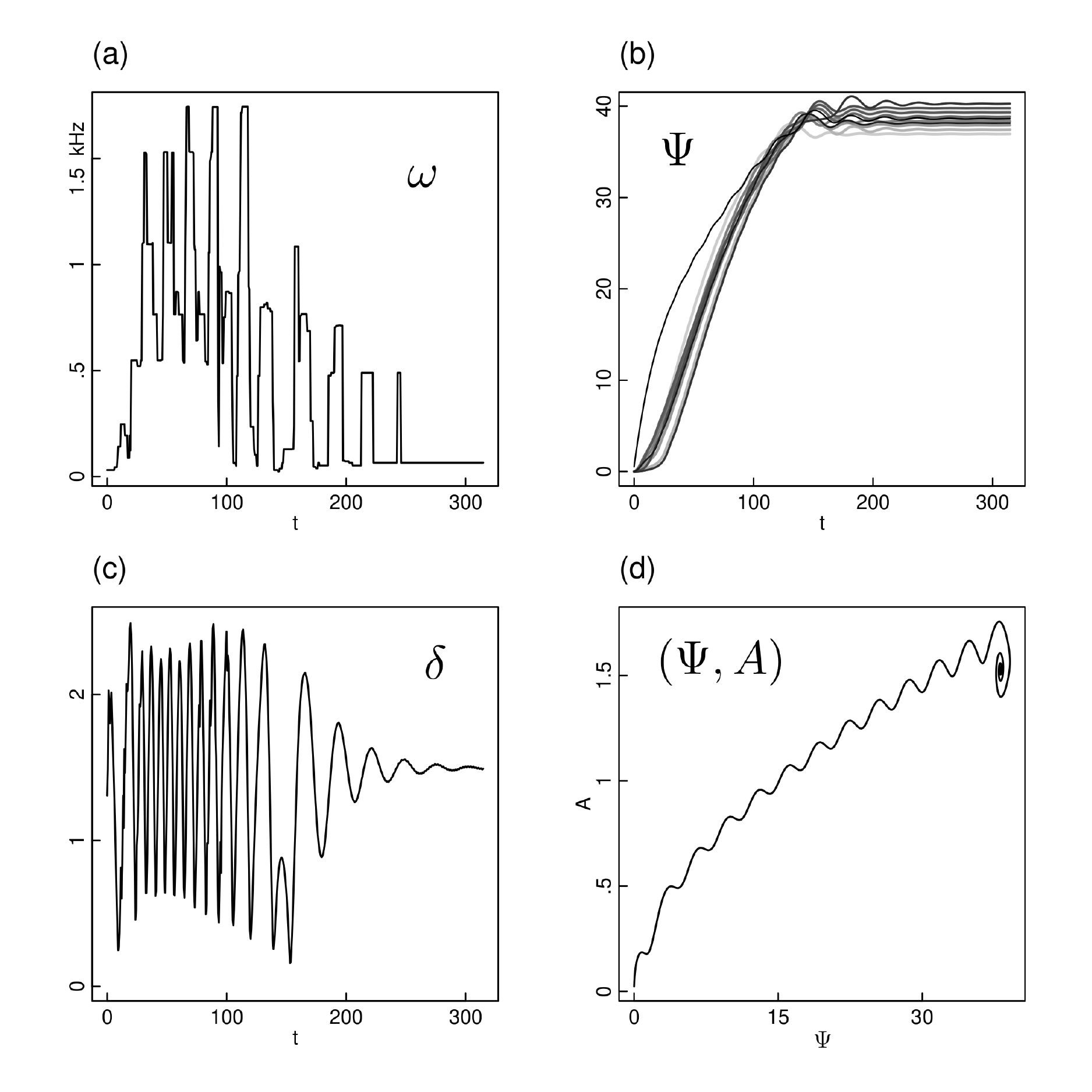}
\par\end{centering}
\caption{Dynamics of the Megaphone system. (\emph{a}) One of the frequency variables
$\omega$ over time in seconds. (\emph{b}) All thirteen slowest variables.
(\emph{c}) Detuning over time, and (\emph{d}) $\psi$ against \emph{A}
showing the nature of the long transient towards a fix point. \label{fig:megaphone-plots}}
\end{figure}

This description still omits a few details but provides the gist of
this particular autonomous system. To summarise, there are the fast
variables (the audio output and oscillator phases), the slower amplitude
envelopes \emph{A}, and the lethargic second order envelope followers
$\psi$. The function $g()$ was designed to produce stepwise changes
at a rate that can be controlled to some extent by setting appropriate
parameter values. Nevertheless, this system goes through an extended
initial transient over a period of a few minutes (depending on sampling
rate and many other parameters) before ending up on a steady state
(see Figure \ref{fig:megaphone-plots}). It is not impossible that there are 
stable periodic oscillatory or chaotic states in parts of the parameter
space, however, it is not very practical to search the space for different
dynamics given the long duration of transients.

Notice also that the discrete output range of the \emph{g} function
means that when it enters a stable attracting state (that is, the
\emph{A}'s end up in a certain order of magnitudes), no small fluctuations
or a gradual approach to the stable state takes place, it ends up
there quite abruptly. Instead of trying to scaffold further layers
of control to add complexity and longevity to the system's errant
dynamics one might consider having a sensor analysing the output and
turning it off as soon as the equilibrium state is reached.

\section*{Concluding remarks}

Autonomous dynamic systems can be used for generative music or algorithmic
composition with systems that integrate all levels, from sound synthesis
to phrase level and formal sections. Monolithic systems of this kind
have channels between their fast and slow subsystems through which
the different levels can influence each other.

Music created with an uncompromising insistence on using the automous
system's output as is, without editing or mixing with other material,
may offer a certain conceptual clarity while also bearing the marks
of dynamic systems. One frequently observed phenomenon is a prolonged initial
transient as the system approaches an attractor. 

Long-lived chaotic transients have been observed in various settings,
e.g. in networks of pulse-coupled oscillators \citep{zumdieck_transient}.
In this type of network the transient length depends on the connectivity
between oscillators. If either a small number of oscillators or most
of the oscillators are connected the transients are short, but at
intermediate degrees of connectivity there can be very long transients.
At intermediate connectivity the average transient length grows exponentially
with the total number of oscillators. When long transients are observed
in algorithmic composition with networks of oscillators and other
signal processing units, these transients could conceivably follow
a similar law of scaling with connectivity and network size. 

Even if the system does not reach an equilibrium state after a prolonged
transient phase, another common observation is that it seems to enter
recognisable patterns after a while. As a composer one is tempted
to compensate for any lack of variety by adding layers of control
to ensure development also on longer time-scales. Since the system's
behaviour may differ dramatically between different positions in its
parameter space as well as initial conditions,
one may need to search for a \textquotedblleft sweet
spot\textquotedblright. The introduction of multiple temporal scales
by explicit design of slow-fast systems, optionally with statistical
feedback, is another convenient way to ensure variation on multiple
levels. And, it should be added, although variation over multiple
time-scales perhaps characterises most music, the deliberate avoidance
of variation on some time-scale might be an interesting avenue to
explore. 

A musical motivation for using autonomous algorithmic composition
systems is to create as much complexity as possible using as simple
means as possible. As with fusion reactors, one hopes, so to speak,
to get more energy out of them than one puts into them to ignite the
process, and perhaps the old saying that fusion energy is always 30
years away also holds for this flavour of algorithmic composition.
Working with closed, deterministic, autonomous systems imposes strict
limits on what is possible, the contours of which are not as easily
seen in open, interactive, and stochastic systems. 

As mentioned previously, there have been efforts to quantify notions
such as complexity, emergence and self-organisation by comparing the
information content at the system's input to that at its output. A
related concept is the Kolmogorov complexity, developed independently
by Kolmogorov, Solomonoff and Chaitin, and also known as algorithmic
complexity \citep{Prokopenko_etal}, which is defined in terms of
universal Turing machines. The Kolmogorov complexity of an object's
description as a text string is defined as the length of the shortest
program that produces that output string. Using the same computer
and programming language, several output strings can be compared to
find out which one is more complex than the other. Still, there may
be no practical way of finding the shortest possible program when
the output is a soundfile consisting of a piece of complex music.
This is equivalent to finding an optimal compression scheme for the soundfile.

Although Kolmogorov complexity is of greater value from a theoretical
than from a practical point of view, it suggests an interesting challenge
for algorithmic composition -- namely, to generate as much complexity
as possible with as little code as possible. Of course it may take
more effort to formulate a concise program than to write a longer
piece of equivalent code. The program's brevity also resembles the
concept of elegance that \citet{Sprott_2010} has applied in his search
for algebraically simple chaotic flows. Elegance, according to Sprott,
is defined as the simplicity of a system of equations where the number
of linear and nonlinear terms are counted; the fewer and simpler the
more elegant the system is. This concept of elegance is obviously
applicable to algorithmic composition using dynamic systems. 

These ideas of simplicity or elegance of program code, while producing
complex output, have been a guiding principle for a collection of
works including those mentioned in the Case study\footnote{Titled \emph{Kolmogorov Variations} or Eleven Hard Pieces. Source code for generating the pieces is available at \url{https://ristoid.net/prog/kolmogorov.html}.}. We have also noted the opposite temptation of 
adding more layers of mechanisms  to increase the complexity of the output. 
Musical complexity and conciseness of code are two goals usually at odds with one another. 

Given that the output is a musical composition, we would like to evaluate
its complexity according to perceptual criteria. There is no single
agreed upon definition of musical complexity, although it may be best
thought of as a multi-dimensional concept. Quantifiable approaches
to measuring musical complexity from audio recordings have been proposed
in music information retrieval, including one that takes structural
change over multiple temporal scales into consideration \citep{Mauch_Levy_2011}.
In the end it is the composer's judgement of the algorithm's output
that matters. 

Preference for musical complexity has been thought to follow an inverted
U-curve; very simple or extremely complex pieces are less liked than
pieces of moderate complexity. Recent research indicates that it is
more revealing to consider two groups of subjects: those who prefer
simplicity and those who prefer complex stimuli \citep{gucluturk_lier}.
These two groups were found to be separated by age, gender and the
personality trait of being susceptible to systemising and analysing.
It was found that the preference for complex stimuli was more common
among men, young subjects and those with high scores of the systemising
quotient, although other factors may contribute. Inasmuch as 
algorithmic composition requires a systemising mentality,
one should not be surprised to find that composers who engage with
this type of systems have a predilection for complex results.

\section*{Acknowledgements }

This project has been realised with funding from The Audio and Visual
Fund, Norway.

\bibliographystyle{apalike}
\bibliography{odelit}

\end{document}